\documentclass[10pt,aps,prb,twocolumn,showpacs]{revtex4}
\usepackage{amsmath,amssymb}
\usepackage{amsfonts}
\usepackage{color}
\usepackage{epsfig}
\usepackage{graphicx}
\usepackage{epstopdf}
\begin{document}

\title{Domain walls and bubble-droplets in immiscible binary Bose gases}
\author{G. Filatrella$^{1}$ }
\author{ Boris A. Malomed$^{2}$ }
\author{ Mario Salerno$^{3}$ }
\affiliation{$^1$ Dipartimento di Scienze e Tecnologie dell'Universit\`a del Sannio,
I-82100 Benevento, Italy}
\affiliation{$^2$ Department of Physical Electronics, School of Electrical Engineering,
Faculty of Engineering, Tel Aviv University, Tel Aviv 69978, Israel}
\affiliation{$^3$ Dipartimento di Fisica "E. R. Caianiello", CNISM and INFN Gruppo
Collegato di Salerno, Universit\`{a} di Salerno, via Giovanni Paolo II
Stecca 8-9, I-84084, Fisciano (SA), Italy.}
\date{\today }

\begin{abstract}
The existence and stability of domain walls (DWs) and bubble-droplet (BD)
states in binary mixtures of quasi-one-dimensional ultracold Bose gases with
inter- and intra-species repulsive interactions is considered. Previously,
DWs were studied by means of coupled systems of Gross-Pitaevskii equations
(GPEs) with cubic terms, which model immiscible binary Bose-Einstein
condensates (BECs). We address immiscible BECs with two- and three-body
repulsive interactions, as well as binary Tonks--Girardeau (TG) gases, using
systems of GPEs with cubic and quintic nonlinearities for the binary BEC,
and coupled nonlinear Schr\"{o}dinger equations with quintic terms
for the TG gases. Exact DW\ solutions are found for the symmetric BEC
mixture, with equal intra-species scattering lengths. Stable asymmetric DWs
in the BEC mixtures with dissimilar interactions in the two components, as
well as of symmetric and asymmetric DWs in the binary TG gas, are found by
means of numerical and approximate analytical methods. In the BEC system,
DWs can be easily put in motion by phase imprinting. Combining a DW and
anti-DW on a ring, we construct BD states for both the BEC and TG models.
These consist of a dark soliton in one component (the 
``bubble"), and a bright soliton (the ``droplet") in the
other. In the BEC system, these composite states are mobile too.
\end{abstract}

\maketitle

\section{Introduction}

Domain walls (DWs) represent the most fundamental type of robust structures
which are supported by immiscibility in a variety of binary physical
systems. Commonly known are DWs in magnetics \cite{magnetic}, ferroelectrics
\cite{electric}, and liquid crystals \cite{liquid}. In those media, the
immiscible species are emulated by different orientations of a vectorial
order parameter. Similarly organized are DWs separating temporal-domain
regions occupied by light waves with orthogonal circular polarizations in
bimodal optical fibers \cite{Haelt,me}. Binary Bose-Einstein condensates
(BEC) formed by immiscible atomic species, or immiscible hyperfine states of
the same atom, make it possible to study the formation of DW structures in
superfluids. These structures were studied theoretically in a number of
different settings \cite{Warsaw}, including the extension to BEC with linear
interconversion between the immiscible components \cite{BEClin-coupling},
condensates with long-range dipole-dipole interactions \cite{DD},
three-component spinor BEC \cite{spinor}, and the BEC effectively
discretized by trapping in a deep optical-lattice potential \cite{Mering}.

In addition to the condensates, degenerate Bose gases have been also
experimentally realized in the Tonks-Girardeau (TG) state \cite{TG}, using
tight quasi-one-dimensional traps \cite{1} (for a review of the TG model see
Ref. \cite{3}). In this context, a macroscopic (mean-field-like) description
of the TG gas was proposed in terms of the quintic nonlinear Schr\"{o}dinger equations NLSE \cite{Kolo} and
adopted for various settings \cite{Kolo}$^{\text{-}}$\cite{7}. In this
framework, it was demonstrated that oscillation frequencies derived from the
fermionic hydrodynamic equations, which emulate the hard-core TG gas, are
very close to their counterparts predicted by the quintic NLSE \cite{12,11}.
The NLSE approach was employed to investigate particular nonlinear features
of TG states, including dark solitons \cite{Kolo,Townes,6}, bright solitons
formed by the long-range dipole-dipole interactions \cite{Fatkh}, etc. Exact
solutions for the ground state in mixtures of TG and Fermi gases have been
found too \cite{Minguzzi}.

On the other hand, the mean-field approach does not apply beyond the
framework of the hydrodynamic regime -- in particular, to problems such as
merger of two gas clouds into one \cite{Girardeau}. Recently, however, the
ground state of the binary TG mixture trapped in the harmonic-oscillator
potential was constructed, using the density-functional theory with the
local-density approximation \cite{density-functional, miscible}. Coupled
quintic NLSEs emerge in this case too (although only for particular
parameter settings), where they were used to investigate mainly miscible
ground states.

One expects that DWs, both in BEC and in TG gases, can
exist only in the case of immiscibility. DWs were indeed observed
experimentally in immiscible binary BECs \cite{BEC-experiment}. DWs have
been also investigated in other physical systems described by systems of
continuous or discrete NLSEs, both conservative and dissipative. In
particular, DWs were found in one-dimensional (1D) Heisenberg ferromagnets
as front patterns, both at the classical \cite{HeisC} and quantum \cite%
{HeisQ} level. Similar structures were also found in a dissipative discrete
NLSE \cite{DW-DNLS} and in coupled Ginzburg-Landau equations modeling
convection patterns in 2D \cite{Alik}. To the best of our knowledge,
however, DWs of immiscible binary gases with quintic interactions, of both
BEC and TG types, have not been investigated yet.

The aim of the present paper is to address the existence and stability of
DWs and bubble-droplet (BD) states in binary mixtures of immiscible BEC and
TG gases. To this end, we introduce a system of nonlinearly-coupled
cubic-quintic (CQ) NLSEs, which describes, in proper situations, both BEC
mixtures and binary TG gases. As concerns the BEC, the quintic
nonlinearities model three-particle collisions in the limit when the related
losses may be neglected \cite{CQ}. In this case, we derive a class of exact
DW solutions, with equal background amplitudes and equal intra-species and
inter-species interactions (symmetric DWs).

The existence and stability of asymmetric DWs in the BEC mixtures with
dissimilar background amplitudes and interactions in the two components, as
well as of symmetric and asymmetric DWs in the binary TG gas, is
demonstrated by means of numerical and approximate analytical methods. It is
shown too that DWs can be easily put in motion by phase imprinting. We also
show that, by combining a DW with an anti-DW on a ring, it is possible to
construct, for both BEC and TG gases, BD excitations, consisting of a dark
(gray) soliton in one component (the ``bubble"), and a
bright soliton (the ``droplet") in the other. In the BEC
system, such composite excitations are found to be mobile as well. Finally,
our estimates suggest that, using the Feshbach-resonance technique to
control the strengths of the inter- and intra-species repulsion in a binary
Bose gas loaded into a quasi-1D trap, the observation of DWs and BDs should
be possible in experiments.

The paper is organized as follows. In Sec. II we introduce the model
equations, for which exact symmetric DWs and the immiscibility condition are
explicitly derived. Section III is focused on symmetric and asymmetric DWs
and BD complexes in the BEC mixture. The existence of asymmetric DWs and
stability of DWs and BDs is numerically investigated, and their mobility is
demonstrated. In Section IV we consider DWs and BDs in binary TG gases for
the symmetric case of equal masses and equal interactions, as well as for
the asymmetric setting. In Section V we summarize the paper and
briefly discuss possible experimental settings.

\section{Model equations and exact DW solutions}

We start with the general system of coupled scaled NLSEs with the CQ
nonlinearity and equal atomic masses of both components:
\begin{eqnarray}
&&i\frac{\partial \psi _{j}}{\partial t}=-\frac{1}{2}\frac{\partial ^{2}\psi
_{j}}{\partial x^{2}}+\left[ \gamma _{j}|\psi _{j}|^{2}+\gamma _{12}|\psi
_{3-j}|^{2}\right] \psi _{j}+  \label{coupledNLSE} \\
&&\left[ \alpha _{j}|\psi _{j}|^{4}+\chi (|\psi _{3-j}|^{4}+2|\psi
_{j}|^{2}|\psi _{3-j}|^{2})\right] \psi _{j},\;\;\;\;\;j=1,2,  \notag
\end{eqnarray}%
where positive real parameters $\gamma _{j},\gamma _{12},\alpha _{j},\chi $
represent the repulsive interactions. We will focus on two physically
relevant cases: (i) $\chi =0$, and (ii) the quintic-only interactions, $%
\gamma _{1,2}=\gamma _{12}=0$. The former case corresponds to the binary BEC
with coefficients $\gamma _{1,2}$ and $\gamma _{12}$ accounting for the
intra-species and inter-species two-body repulsive interactions,
respectively, while $\alpha _{1,2}$ add the repulsive three-body
interactions in each component \cite{CQ}. On the other hand, the
quintic-only version of Eq. (\ref{coupledNLSE}) may be used as the model for
the binary TG gases.

Obviously, Eq. (\ref{coupledNLSE}) admits the Hamiltonian representation, in
the form of $i\partial \psi _{j}/\partial t=\delta H/\delta \psi _{j}^{\ast }
$, with
\begin{eqnarray}
H &=&\int_{-\infty }^{+\infty }\sum_{j=1}^{2}\left\{ \frac{1}{2}\left\vert
\frac{\partial \psi _{j}}{\partial x}\right\vert ^{2}+\frac{\gamma _{j}}{2}%
|\psi _{j}|^{4}+\frac{\alpha _{j}}{3}|\psi _{j}|^{6}+\right.   \notag \\
&&\left. \frac{\gamma _{12}}{2}|\psi _{j}|^{2}|\psi _{3-j}|^{2}+\chi |\psi
_{j}|^{4}|\psi _{3-j}|^{2}\right\} dx.  \label{Hamiltonian}
\end{eqnarray}%
In particular, this representation explains ratio $1:2$ of coefficients in
front of the terms multiplied by $\chi $ in Eq. (\ref{coupledNLSE}), which
are derived from terms $\chi \left( \left\vert \psi _{1}\right\vert
^{4}\left\vert \psi _{2}\right\vert ^{2}+\left\vert \psi _{2}\right\vert
^{4}\left\vert \psi _{1}\right\vert ^{2}\right) $ in the Hamiltonian density
of Eq. (\ref{Hamiltonian}). In addition to $H$, the system preserves the
norm (scaled number of the atoms) of each component,
\begin{equation}
N_{j}=\int_{-\infty }^{+\infty }|\psi _{j}|^{2}dx,  \label{N}
\end{equation}%
and the total momentum, $P=i\int_{-\infty }^{+\infty }\sum_{j=1}^{2}\psi
_{j}\left( \partial \psi _{j}^{\ast }/\partial x\right) dx$.

Due to the repulsive nature of the nonlinearity, one may expect that, with
the increase of the constants accounting for the interactions between the
components, $\gamma _{12}$ and/or $\chi $, the binary system becomes
immiscible, building DWs as interfaces between domains filled by different
components. To address this point, we are first looking for particular \emph{%
exact} DW solutions to Eq. (\ref{coupledNLSE}). In the case of the coupled
stationary NLSEs with the cubic nonlinearity, which corresponds to Eq. (\ref%
{coupledNLSE}) with $\alpha _{1,2}=\chi =0$, an exact DW solution was found
in Ref. \cite{Alik}, imposing a special restriction on the cubic
coefficients, $\gamma _{1}=\gamma _{2}=\gamma _{12}/3$. For the
single-component NLSE with the CQ nonlinearity, an exact solution,
describing a transient layer between zero and constant-amplitude states (a
variety of DW), was found in Ref. \cite{Zeev}. In the present context, we
are looking for exact DW solutions employing an ansatz suggested by the
latter solution:
\begin{equation}
\psi _{1}(x,t)=\frac{A_{1}\,e^{-i\mu _{1}t}}{\sqrt{1+e^{\lambda x}}}%
,\;\;\psi _{2}(x,t)=\frac{A_{2}\,e^{-i\mu _{2}t}}{\sqrt{1+e^{-\lambda x}}},
\label{ansatz}
\end{equation}%
with background amplitudes $A_{1,2}$, chemical potentials$\;\mu _{1,2}$, and
parameter $\lambda $ defining the DW width, $W\sim 1/\lambda $.

Substituting this ansatz into Eq. (\ref{coupledNLSE}), one arrives at a
system of algebraic equations:
\begin{eqnarray}
&&4(A_{j}^{2}\gamma _{j}+A_{3-j}^{2}\gamma _{12}+2\chi
A_{j}^{2}A_{3-j}^{2})-8\mu _{j}+\lambda ^{2}=0,  \notag \\
&&8A_{j}^{2}(\gamma _{12}+\chi A_{j}^{2})-8\mu _{j}-\lambda ^{2}=0,
\label{sol-param} \\
&&A_{j}^{2}(\gamma _{j}+\alpha _{j}A_{j}^{2})-\mu
_{j}=0,\;\;\;\;\;\;\;\;\;\;\;\;j=1,2.  \notag
\end{eqnarray}%
One can readily check that for symmetric intra-species interactions, i.e.,
\begin{equation}
\gamma _{1}=\gamma _{2}\equiv \gamma ,\alpha _{1}=\alpha _{2}\equiv \alpha ,
\label{alpha}
\end{equation}
and equal amplitudes (the symmetric DW), Eq. (\ref{sol-param}) admits a
nontrivial solution with $\mu _{1}=\mu _{2}\equiv \mu $, $A_{1}=A_{2}\equiv
A $, and
\begin{eqnarray}
&&A^{2}=\frac{3}{4}\frac{\gamma _{12}-\gamma }{\alpha -\chi },  \notag \\
&&\lambda =\pm (\gamma _{12}-\gamma )\sqrt{\frac{3}{2(\alpha -\chi )}},
\label{exact} \\
&&\mu =\frac{3}{16}\frac{\gamma _{12}-\gamma }{(\alpha -\chi )^{2}}(\alpha
\gamma +3\alpha \gamma _{12}-4\gamma \chi ).  \notag
\end{eqnarray}%
From this we conclude that symmetric DWs are generic solutions, provided
that conditions $\alpha >\chi $ and $\gamma _{12}>\gamma $ are satisfied.
Note that in the decoupling limit, $\gamma _{12}\rightarrow 0,\chi
\rightarrow 0$, and for the symmetric interactions, Eq. (\ref{coupledNLSE})
reduces to a single-component CQ NLSE. In this case, the exact DW reproduces
the one found in Ref. \cite{Zeev} for the particular case of $\gamma
=-2,\alpha =1$.

\subsection{The immiscibility condition and asymptotic relations}

It is interesting to relate the existence of generic DW solutions to the
immiscibility in the two-component NLSE system (\ref{coupledNLSE}). Due to
the repulsive nature of all the interactions and the absence of any trapping
potential, the ground state must be obviously spatially uniform in the
miscible case, hence the DW cannot exist in the miscible system. In the case
of immiscibility, generic stable DW can exist if its free energy is lower
than the one of the corresponding uniformly mixed state. Then, the
immiscibility condition is written in terms of the free energy, $F=H[\psi
_{1},\psi _{2}]-\sum_{j}\mu _{j}N_{j}$, as
\begin{equation}
F_{\mathrm{DW}}-F_{\mathrm{UB}}\leq 0,  \label{DWimm}
\end{equation}%
where $F_{\mathrm{DW}}-F_{\mathrm{UB}}$ is the difference in the free energies
between the DW and of the corresponding uniform background (each free energy
diverges in the infinite system, but the difference is finite).
\begin{figure}[t]
\includegraphics[width=0.4\textwidth]{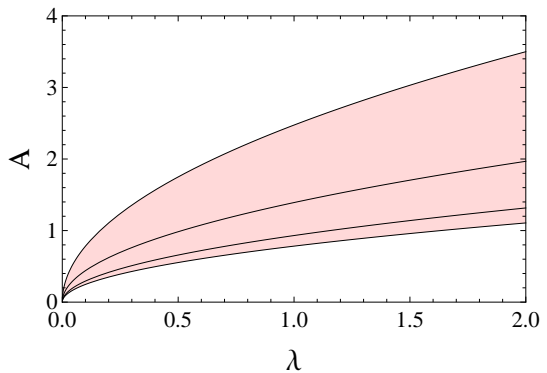}
\caption{(Color online) The existence region (shaded) for families of
stationary exact domain-wall solutions given by Eqs. (\protect\ref{ansatz})
and (\protect\ref{exact}) under condition (\protect\ref{alpha}), with the
quintic coefficient, $\protect\alpha $, varying in the interval of $[0.01,1]$%
. Continuous curves from top to bottom refer to $\protect\alpha %
=0.01,0.1,0.5,1.0$, respectively.}
\label{fig1}
\end{figure}

Defining background amplitudes $A_{j},\;j=1,2$, of the two DW components as
in ansatz (\ref{ansatz}),
\begin{equation}
\left\{
\begin{array}{c}
|\psi _{1}(x=-\infty )|^{2}=A_{1}^{2} \\
|\psi _{2}(x=-\infty )|^{2}=0%
\end{array}%
\right\} ,~\left\{
\begin{array}{c}
|\psi _{1}(x=+\infty )|^{2}=0 \\
|\psi _{2}(x=+\infty )|^{2}=A_{2}^{2}%
\end{array}%
\right\}  \label{asympt}
\end{equation}%
%
(the corresponding densities in the uniformly mixed state obviously being $%
|\psi _{j}|^{2}=A_{j}^{2}/2$), one can readily write the immiscibility
condition (\ref{DWimm}), with the aid of Eq. (\ref{Hamiltonian}), in the
explicit form:
\begin{equation}
\sum\nolimits_{j=1,2}(\gamma _{j}A_{j}^{4}+\alpha _{j}A_{j}^{6})\leq \left[
2\gamma _{12}+\chi (A_{1}^{2}+A_{2}^{2})\right] A_{1}^{2}A_{2}^{2},
\label{immC}
\end{equation}%
which is exact for the infinite domain, and approximated for a finite domain
of length $L$, the error coming from the gradient-energy terms being
estimated as $\sim 1/L$. For the exact symmetric DW solution given by Eq. (%
\ref{exact}), condition (\ref{DWimm}) is always satisfied. Indeed, from Eq. (%
\ref{exact}) we obtain
\begin{equation*}
A^{2}=\frac{3}{4}\frac{\gamma _{12}-\gamma }{\alpha -\chi }<\frac{\gamma
_{12}-\gamma }{\alpha -\chi },
\end{equation*}%
this being in agreement with Eq. (\ref{immC}), taking Eq. (\ref{alpha}) into
regard. It is also relevant to note that the equality of the free-energy
densities at $x=\pm \infty $ implies the following relation between the
chemical potentials and asymptotic densities of the numbers of atoms, fixed
as per Eq.(\ref{asympt}):
\begin{equation}
\sum\nolimits_{j=1,2}(-1)^{j}\left[ \mu _{j}-\frac{1}{2}\gamma
_{j}A_{j}^{2}-\frac{\alpha _{j}}{3}A_{j}^{4}\right] A_{j}^{2}=0.  \label{amu}
\end{equation}

For $\chi =\alpha $ and $\gamma =\gamma _{12}=0$, which corresponds to the
two-component TG\ gas described by the coupled quintic equations, exact
solution (\ref{exact}) degenerates into a uniform one. 
Different solutions for this case are given below.

\section{Repulsive BEC mixtures with cubic and quintic interactions}

For the binary BEC mixture, we consider a reduced form of the NLSE system (%
\ref{coupledNLSE}), in which the inter-species repulsion is accounted for by
the cubic terms, while the quintic ones contribute solely to the
self-repulsion ($\chi =0$):
\begin{eqnarray}
i\frac{\partial \psi _{1}}{\partial t} &=&-\frac{1}{2}\frac{\partial
^{2}\psi _{1}}{\partial x^{2}}+(\gamma _{1}|\psi _{1}|^{2}+\gamma _{12}|\psi
_{2}|^{2}+\alpha _{1}|\psi _{1}|^{4})\psi _{1},\;\;\;\;\;\;\;  \notag \\
i\frac{\partial \psi _{2}}{\partial t} &=&-\frac{1}{2}\frac{\partial
^{2}\psi _{2}}{\partial x^{2}}+(\gamma _{2}|\psi _{2}|^{2}+\gamma _{12}|\psi
_{1}|^{2}+\alpha _{2}|\psi _{2}|^{4})\psi _{2},\;\;\;\;\;  \label{GPE}
\end{eqnarray}%
\begin{figure}[t]
\includegraphics[width=0.35\textwidth]{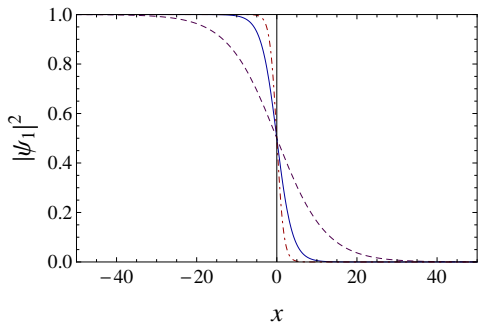}
\caption{(Color online) Typical profiles of stationary domain walls in the
binary immiscible BEC generated by Eq. (\protect\ref{GPE}) with values of
the quintic coefficient $\protect\alpha =0.01,0.1,$ and $0.05$ (purple
dashed, blue continuous, and red dotted curves, respectively). The
parameters are taken as per Eq. (\protect\ref{A=1}), so as to have $A=1$ in
Eq. (\protect\ref{exact}). We only show the profiles of the first component, the other one being produced by specular reflection with respect to the vertical axis. }
\label{fig2}
\end{figure}
\begin{figure}[t]
\caption{A moving domain wall produced by phase imprinting with initial
velocity $v=0.8$, see the text.}
\label{fig3}
\includegraphics[width=0.35\textwidth]{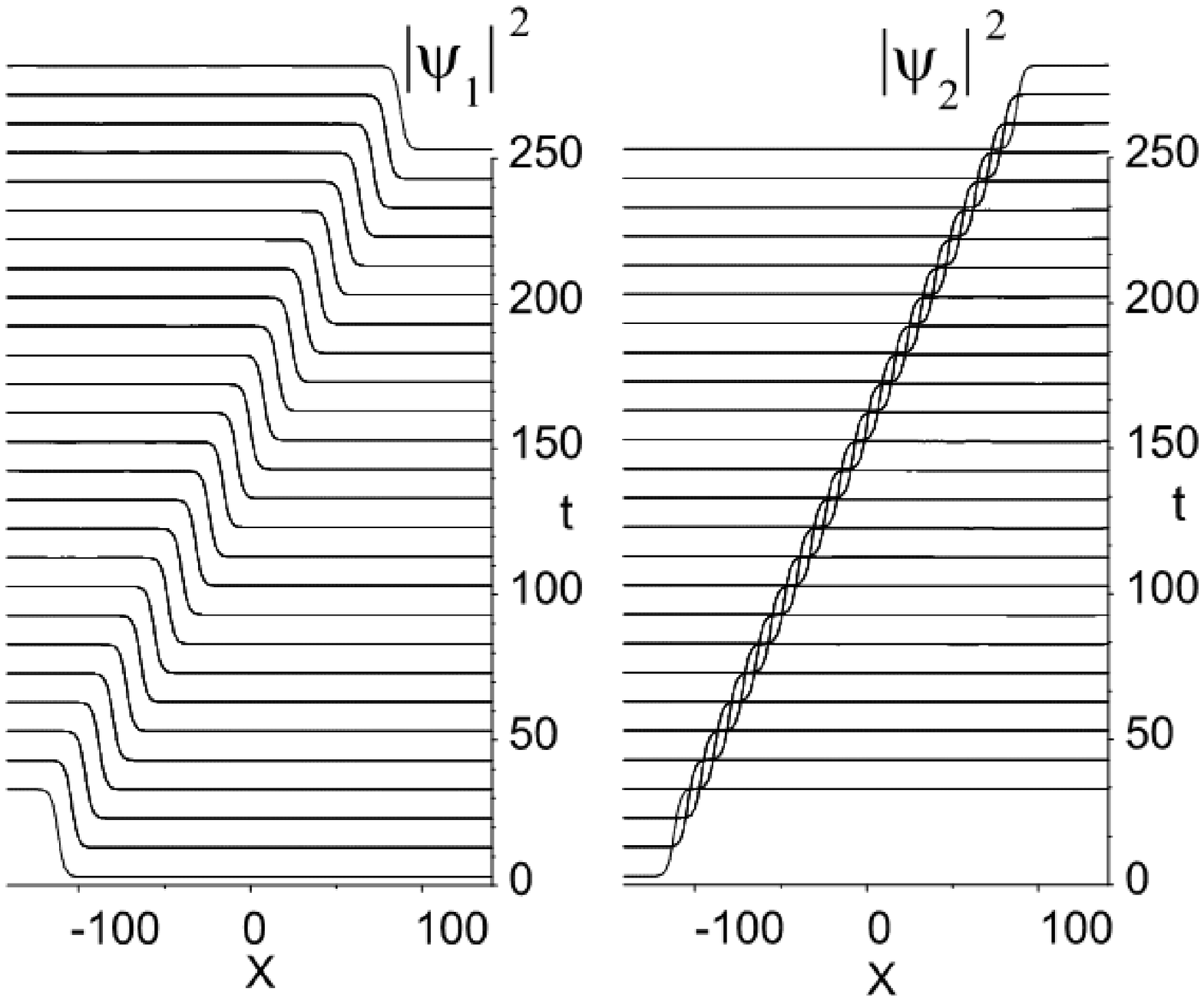} \vskip -3cm
\end{figure}

\subsection{Symmetric and asymmetric DWs in the infinite domain}

For the symmetric system, as defined in Eq. (\ref{alpha}), parameters of the
exact DW solution given by Eqs. (\ref{ansatz}) and (\ref{exact}) with $\chi
=0$ are
\begin{equation}
\mu =\frac{\gamma +3\gamma _{12}}{4}A^{2},\;\;\lambda =\sqrt{\frac{8\alpha }{%
3}}A^{2},\;\;A^{2}=\frac{3}{4}\frac{\gamma _{12}-\gamma }{\alpha }.
\label{BECparam}
\end{equation}%
In Fig. \ref{fig1} we show the existence region of this solution in the
plane of $(A,\lambda )$ (the amplitude and inverse width of the DW), as the
three-body coefficient, $\alpha $, varies in the interval of $[0.01,1]$.
Typical profiles of the corresponding stationary domain walls for different
values of the three-body interaction parameter $\alpha $ and the background
amplitude fixed to be $A=1$, i.e.,
\begin{equation}
\gamma _{12}=\gamma +(4/3)\alpha ,  \label{A=1}
\end{equation}%
as it follows from Eqs. (\ref{ansatz}) and (\ref{sol-param}), are depicted
in Fig. \ref{fig2}. In this case, as seen from Eq. (\ref{BECparam}), the
inverse width of the DW is determined solely by the quintic parameter, $%
\lambda =\sqrt{8\alpha /3}$. 

As said above, for the exact DW solutions the immiscibility condition (\ref%
{immC}) always holds, hence the stability of these solution is expected.
This has been checked by simulations of Eq. (\ref{GPE}), with small noise
added to the initial DW profile (not shown here in detail). We have found
that the DW remains stable also when it is put in motion by means of the
standard phase-imprinting method, i.e., multiplying the quiescent DW by $%
\exp \left( ivx\right) $, as shown in Fig. \ref{fig3}.

\begin{figure}[tbph]
\centerline{\includegraphics[width=0.27\textwidth]{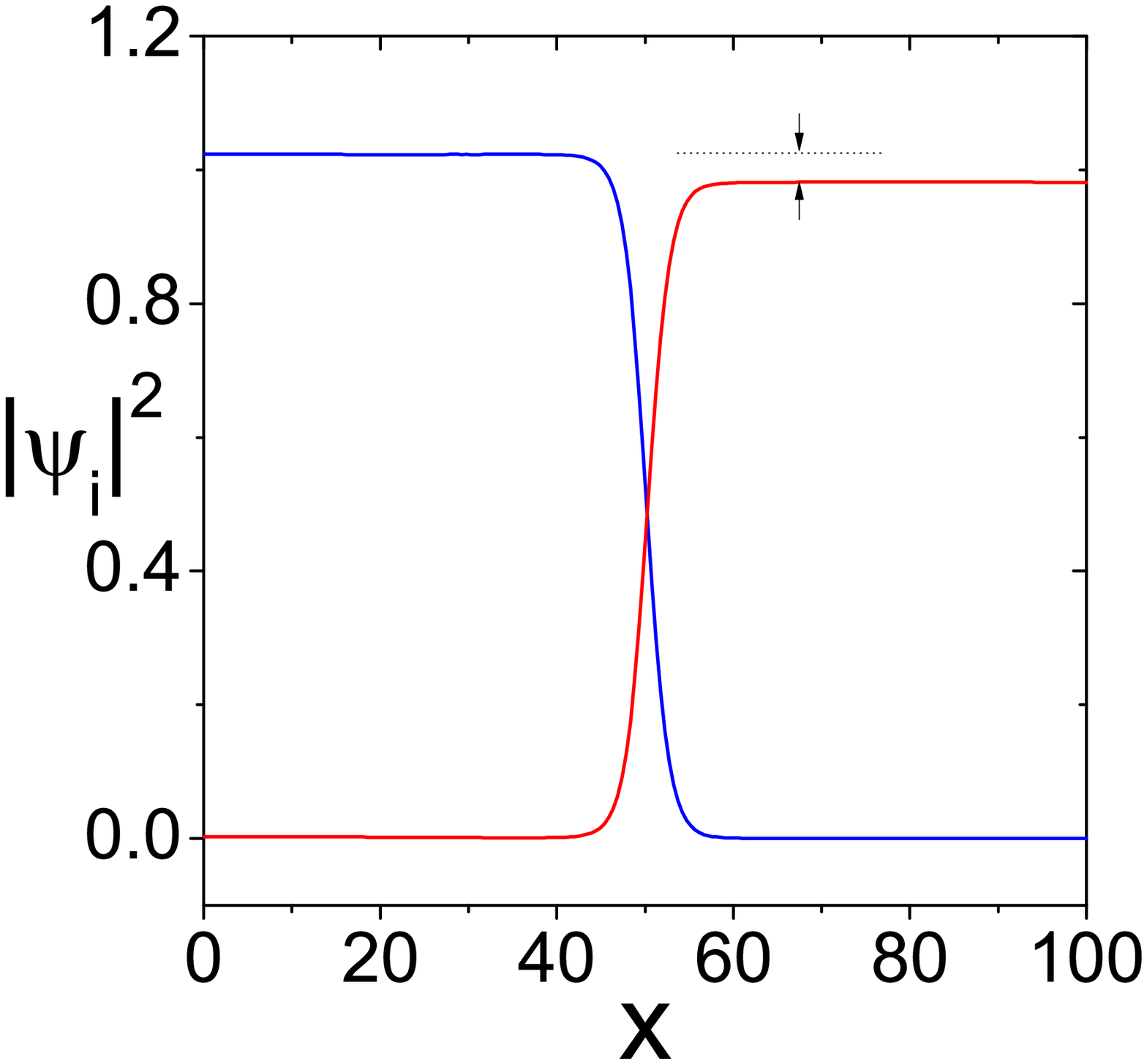}
\hskip -.5cm
\includegraphics[width=0.27\textwidth]{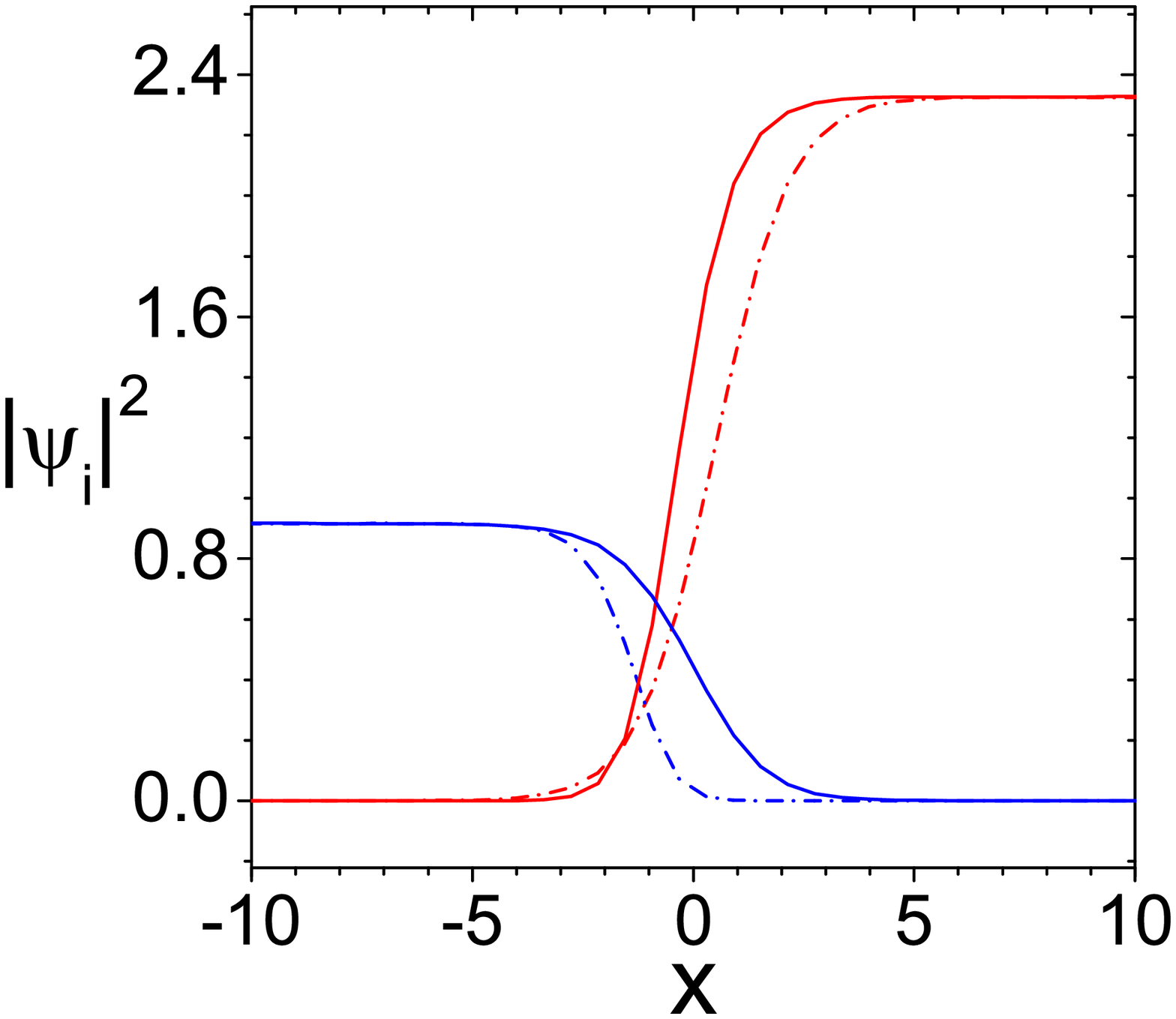}} \vskip -2.cm
\caption{(Color online) The left panel: Typical density profiles of an asymmetric
stationary domain wall in the binary immiscible BEC with $\protect\gamma %
_{1}=1,\protect\gamma _{2}=0.9,\protect\gamma _{12}=1.08$ and equal
coefficients of the quintic nonlinearity, $\protect\alpha _{1}=\protect%
\alpha _{2}=0.1$. The short horizontal line with the vertical arrow shows
the mismatch between the left and right background levels. The right panel:
The same as in the left panel, but for $\protect\gamma _{1}=1$, $\protect%
\gamma _{2}=0$, $\protect\alpha _{1}=0$, $\protect\alpha _{2}=0.05$ in Eq. (%
\protect\ref{GPE}), and different strengths of the interspecies cubic
interaction: $\protect\gamma _{12}=0.5$ (continuous curves), $\protect\gamma %
_{12}=1.1$ (dash-dotted curves) and asymptotic (left and right) densities $%
A_{1}=0.916$, $A_{2}=2.325$. In both panels, the blue and red lines depict
profiles of the first and second components, respectively.}
\label{fig3bis}
\end{figure}

The symmetry restrictions, $\gamma _{1}=\gamma _{2}$, $\alpha _{1}=\alpha
_{2}$, adopted above for obtaining the exact DW solution, is not a
limitation for the existence of DWs in real condensates. Indeed, it is
possible to demonstrate that DWs are generic states in immiscible
two-component BECs with repulsive interactions, including asymmetric
settings with unequal background densities in the two components at $x=\pm
\infty $. A numerically found example of such a stable asymmetric DW is
depicted in the left panel of Fig. \ref{fig3bis}, see also Section V.

Another relevant physical situation, in which strongly asymmetric DW
solutions naturally appear, is the one with the two-body and three-body
intra-species interactions acting only in the first and second components,
respectively, the coupling being accounted for by two-body interactions.
This corresponds to $\gamma _{2}=0,\alpha _{1}=0$ in Eq. (\ref{GPE}), which
can be experimentally implemented by enhancing the three-body interaction in
one component via Efimov states \cite{Efimov}, simultaneously tuning the
two-body scattering length in the same component to zero by means of the
Feshbach resonances. For this case, typical asymmetric DWs are displayed in
the right panel of Fig. \ref{fig3bis}. One can check, using parameter values
given in the caption to the figure, that the respective immiscibility
condition [setting $\chi =0,\alpha _{1}=0,\gamma _{2}=0$ in Eqs. (\ref{immC}%
)] for these solutions,%
\begin{equation}
\gamma _{1}A_{1}^{4}+\alpha _{2}A_{2}^{6}-2\gamma
_{12}A_{1}^{2}A_{2}^{2}\leq 0,  \label{imm1}
\end{equation}%
is satisfied; hence, despite the large mismatch between the backgrounds, one
can expect this asymmetric DW to be stable. This was confirmed by direct
simulations of Eq. (\ref{GPE}) (not shown here).
The respective relation between asymptotic
amplitudes $A_{j}$ and chemical potentials [setting $\chi =0,\alpha
_{1}=0,\gamma _{2}=0$ in Eq. (\ref{amu})],
\begin{equation}
\mu _{1}A_{1}^{2}-\frac{\gamma _{1}}{2}a_{1}^{4}=\mu _{2}A_{2}^{2}-\frac{%
\alpha _{2}}{3}a_{1}^{6},  \label{asy1}
\end{equation}%
was also confirmed numerically. Note that, while Eq. (\ref{asy1}) does not
contain $\gamma _{12}$, the immiscibility condition (\ref{imm1})\ explicitly
depends on it. From the experimental point of view, this suggests to
control the immiscibility condition by keeping parameters $\gamma
_{j},\alpha _{j},\;j=1,2,$ fixed and changing the inter-species two-body
scattering length by means of the Feshbach resonance. The variation of $%
\gamma _{12}$ mainly affects the shape of the DW interface, while the
asymptotic values $A_{j}$ remain unaltered. It is easy to find, from Eq. (%
\ref{imm1}), the critical value at which the mixture becomes miscible in
this case:
\begin{equation}
\tilde{\gamma}_{12}=\frac{\gamma _{1}A_{1}^{4}+\alpha _{2}A_{2}^{6}}{%
2A_{1}^{2}A_{2}^{2}}.  \label{imm-thre}
\end{equation}%
This prediction will be numerically checked in the next subsection (see Fig. %
\ref{fig5ter} below for DWs in finite-length rings).
\begin{figure}[tbph]
\includegraphics[width=0.35\textwidth]{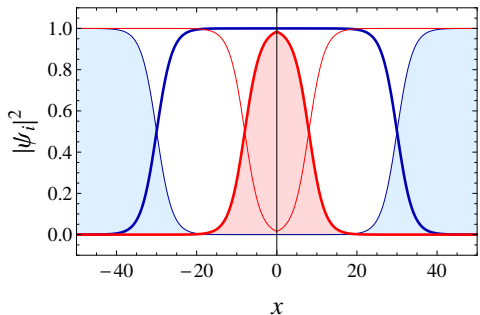}
\caption{(Color online) The DW-anti-DW state in Eq.(\ref{ansatzDB}) of the repulsive binary BEC
with $\protect\alpha =0.1,A=1$, $\lambda =2\sqrt{2 \alpha /3}$ and separation $x_{0}=30$ (blue lines) or $x_{0}=8$ (red lines). 
Thin and thick lines refer to the first and second
components, respectively. Filled blue and red regions depict, severally, the
bubble and droplet parts of the patterns with $x_{0}=30$ and $x_{0}=8$ (they
are shown for different values of $x_{0}$ to avoid confusing overlap between
them). }
\label{fig4}
\end{figure}
%

\subsection{BEC mixtures on a ring: BD states}

The DW solutions considered above refer to infinite domains. DWs in
finite-length ring regions, subject to periodic boundary conditions (b.c.),
are relevant for the realization of the BEC in toroidal traps \cite{ring},
including binary condensates \cite{ring-binary}. Such patterns can be
constructed by combining a DW and the respective anti-DW on the ring, to
satisfy the b.c. More precisely, we consider trial solutions of the type
\begin{eqnarray}
\label{ansatzDB}
\psi _{1}(x,t) &=&\left\{
\begin{array}{ll}
\frac{Ae^{-i\mu t}}{\sqrt{1+e^{\lambda (x+x_{0})}}} & \mbox{at $x \le 0$} \\
\frac{Ae^{-i\mu t}}{\sqrt{1+e^{-\lambda (x-x_{0})}}} & \mbox{at $x > 0$}%
\end{array}%
,\right.  \notag \\
&&  \label{dw-adw} \\
\psi _{2}(x,t) &=&\left\{
\begin{array}{ll}
\frac{Ae^{-i\mu t}}{\sqrt{1+e^{-\lambda (x+x_{0})}}} & \mbox{at $x \le 0$}
\\
\frac{Ae^{-i\mu t}}{\sqrt{1+e^{\lambda (x-x_{0})}}} & \mbox{at $x > 0$}%
\end{array}%
,\right.  \notag
\end{eqnarray}%
where $4x_{0}$ is the perimeter of the underlying ring, and $2x_{0}$ is
separation between the DW and anti-DW placed at diametrically opposite
positions.
\begin{figure}[tbph]
\caption{A moving bubble-droplet solution obtained by the phase imprinting,
with initial velocity $v=1.2$ velocity, onto the stationary profiles shown
in Fig. \protect\ref{fig4} for $x_{0}=8$.}
\vskip 1cm
\label{fig5}
\includegraphics[width=0.35\textwidth]{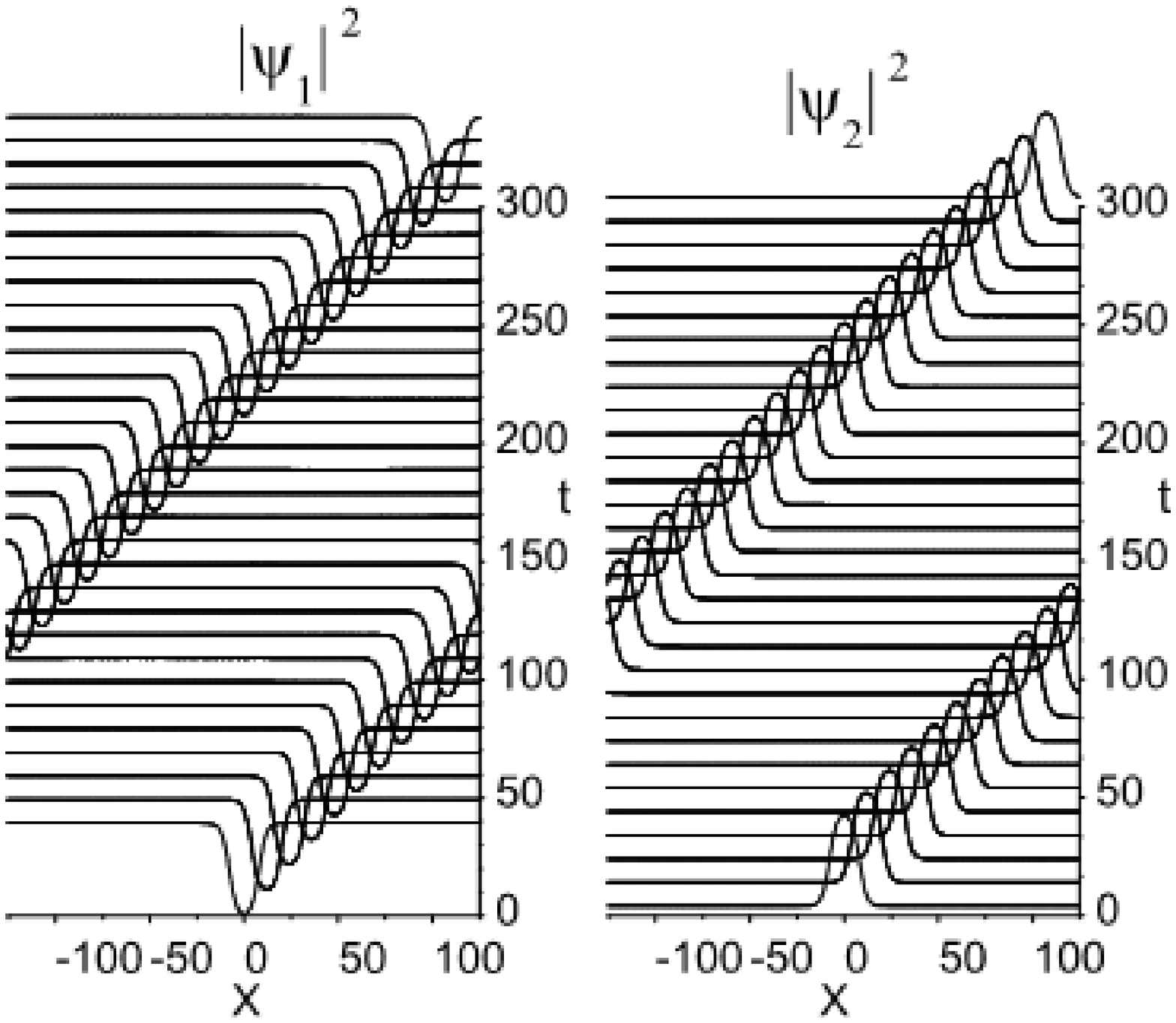} \vskip -0cm
\end{figure}

Typical profiles of such numerically found DW-anti-DWs patterns are
displayed in Fig. \ref{fig4} for two different values of $x_{0}$. One can
expect that ansatz (\ref{dw-adw}) is a virtually exact solution when DW and
anti-DW are well separated, e.g. $x_{0}\gg \lambda ^{-1}$. This is indeed
what one observes from the numerical solutions. More remarkable is the fact
that the ansatz provides an almost exact solution even when the DW and
anti-DW are relatively close to each other, as one can see in Fig. \ref{fig4}%
. In this case, the complex composed of the DW and anti-DW profiles may be
considered as a \textit{bubble} (sort of a dark soliton \cite{Barash}) in
field $\psi _{2}$ coupled to a localized bright profile (\textit{droplet})
of field $\psi _{1}$, which we refer to as BDs. The stability of these
complexes was verified by direct simulations of their perturbed simulations
in the framework of Eq. (\ref{GPE}) (not shown here in detail). 
\begin{figure}[tbph]
\includegraphics[width=0.35\textwidth]{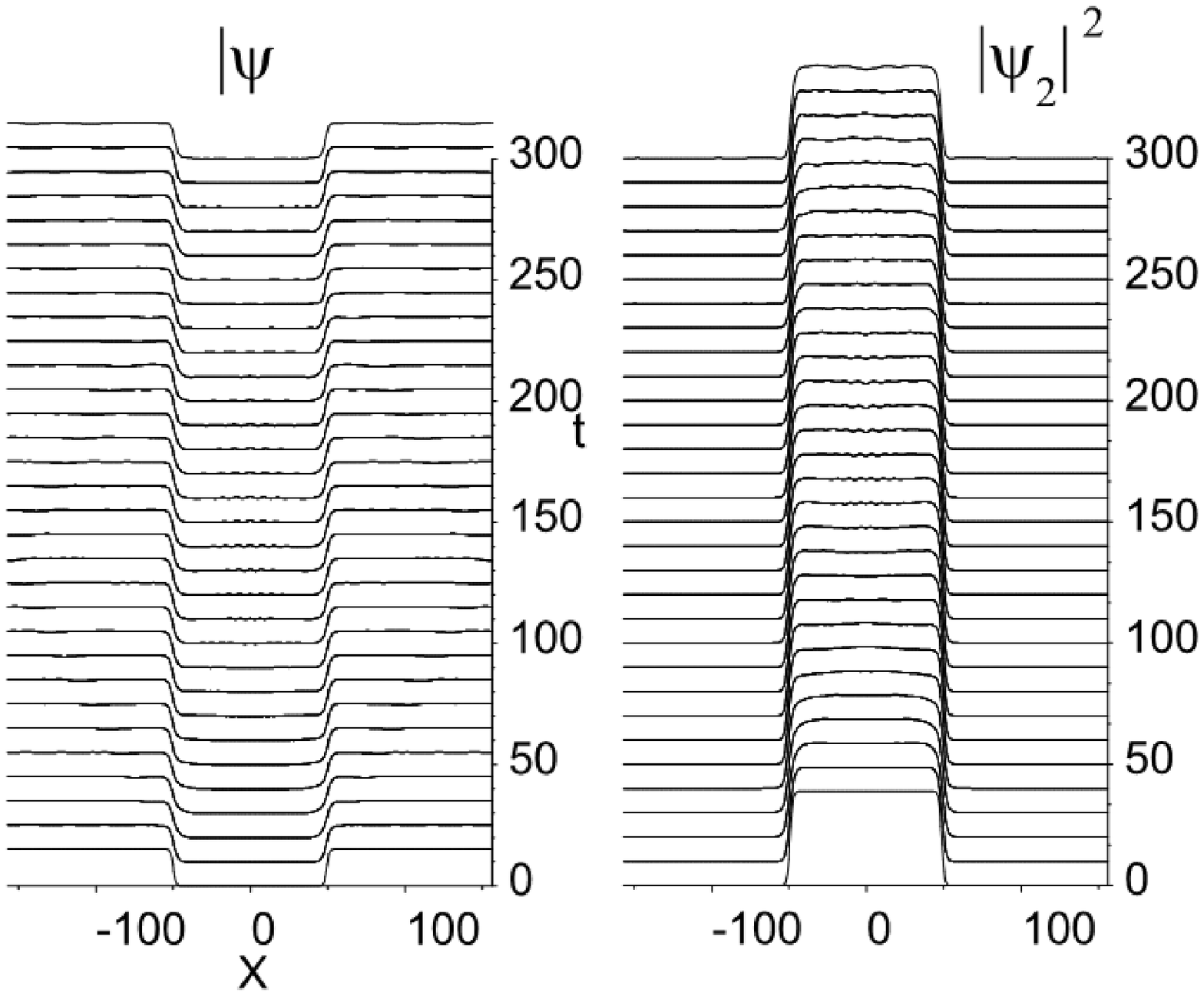} \vskip -3.6cm %
\includegraphics[width=0.35\textwidth]{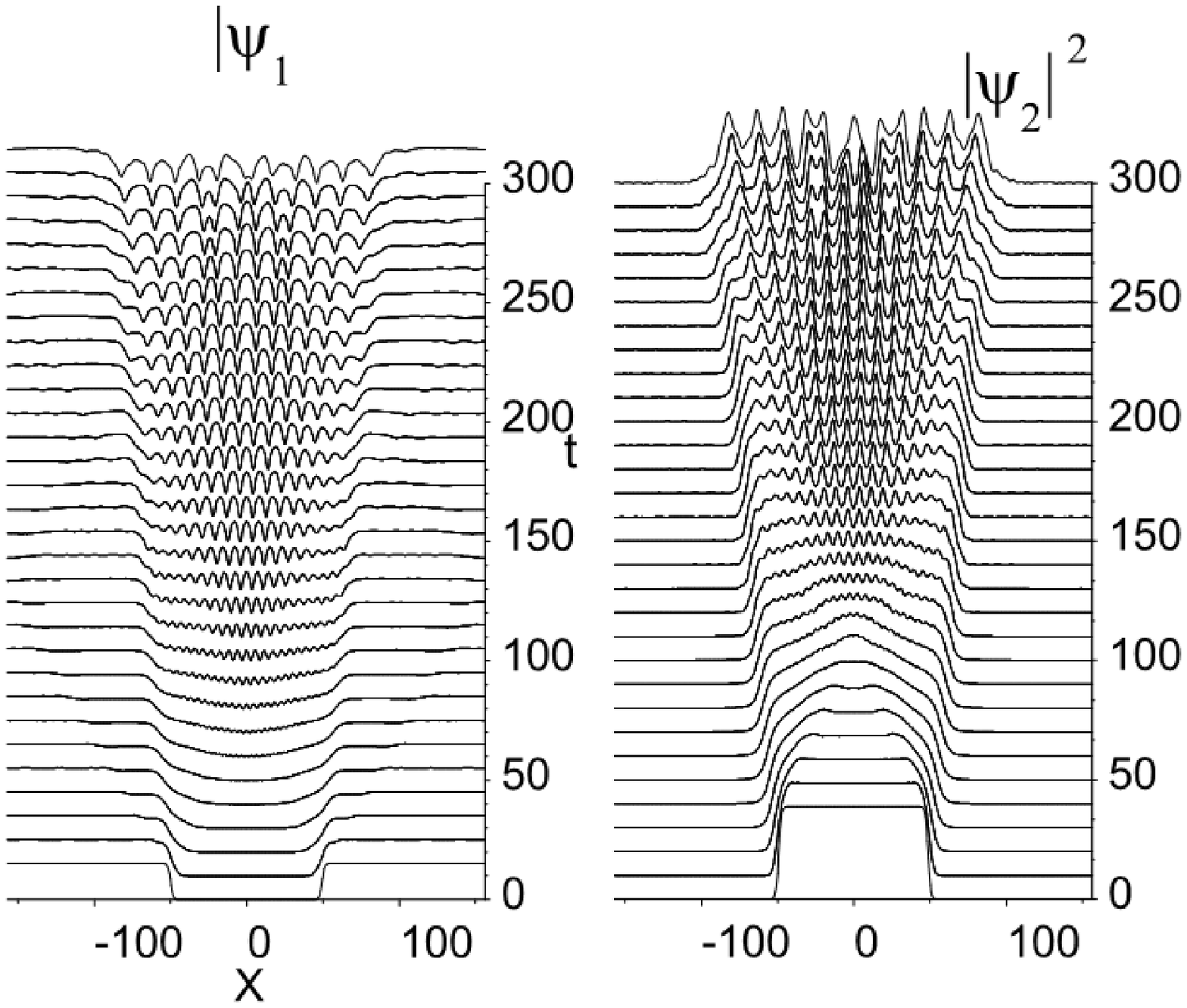} \vskip -3cm
\caption{Top panels: The evolution of densities of the first and second
(left and right columns) components of the BD state for the same parameters
as in the right panel of Fig. \protect\ref{fig3bis}, except for $\protect%
\gamma _{12}=0.4$, which is taken above the miscibility threshold. Bottom
panels: The same as above, but for $\protect\gamma _{12}=0.3$, taken just
below the miscibility threshold.}
\label{fig5ter}
\end{figure}

The possibility to set a DW in motion by phase-imprinting a velocity onto it
is relevant to the BD states too. This is shown in Fig. \ref{fig5}, where
simulations of Eq. (\ref{GPE}) for moving DBs are reported. This figure
makes it clear that the BD structures are very robust ones not only in the
stationary form, but also when they are set in motion, thanks to the
repulsive character of all the interactions.

BD solutions can be constructed as well by combining an asymmetric DW with
the corresponding anti-DW on the ring (not shown here, as their shape is
quite obvious). 
The asymmetric complexes are stable too. 

It is also interesting to check, in terms of the BD solutions, the
miscibility condition discussed above. For the parameter values given in the
caption of Fig. \ref{fig3bis} (the left panel), the miscibility threshold is
predicted by Eq. (\ref{imm-thre}) to occur at $\tilde{\gamma}_{12}\approx
0.344$. In Fig. \ref{fig5ter} we display the evolution of the density
profiles produced by the simulations of Eq. (\ref{GPE}) for $\gamma _{12}$
taken just above and below the threshold, i.e., in the regions of weak
immiscibility and miscibility respectively, taking, as initial conditions, a
BD profile constructed from the asymmetric DW corresponding to the
continuous curves in the left panel of Fig. \ref{fig3bis}. It is seen that,
for $\gamma _{12}=0.4$, the BD solution (and its DW and anti-DW
constituents) quickly adapt, by emitting small-amplitude matter-waves, to
the new value of the inter-species interaction, keeping the immiscibility,
while for $\gamma _{12}=0.3$ the mixing of the two components sets in,
generating strong density waves in the two components.

\section{Domain walls and bubble-drop complexes in binary Tonks-Girardeau
gases}

As said in the Introduction, the NLSE with the quintic nonlinearity emerges
in connection to the ground-state properties of TG gases, both for the
single-component ones \cite{Kolo,5,12,Townes,7} and binary mixtures \cite%
{density-functional, miscible}. In this Section we investigate the existence
of DW states at the interface of two interacting TG gases by means of the
NLSE system (\ref{coupledNLSE}) with only quintic terms included. The
respective equations for stationary states follow from Eqs. (\ref
{coupledNLSE}), substituting $\psi _{1,2}\left( x,t\right) =\exp \left(
-i\mu _{1,2}t\right) \phi _{1,2}(x)$:
\begin{gather}
\left[ -\frac{\hbar ^{2}}{2m_{j}}\frac{d^{2}}{dx^{2}}+\alpha _{j}|\phi
_{j}|^{4}+\right.  \notag \\
 \left.  \phantom{\frac{d^2}{m_j}}  \chi (|\phi _{3-j}|^{4}+2|\phi _{j}|^{2}|\phi _{3-j}|^{2})  \right]
\phi _{j}=\mu _{j}\phi _{j},  \label{coupledTG}
\end{gather}
with $m_{1,2}$ being the atomic masses of the two bosonic species.

For equal masses $m_{1}=m_{2}=m$ and fully symmetric interactions, $\alpha
_{1}=\alpha _{2}=\chi $, this model for the TG mixture can be justified in
terms of the density-functional theory, as discussed in Ref. \cite{miscible}
. In this respect, we recall that the 1D quantum model of two boson species
interacting via the hard-core repulsion is exactly solvable by means of the
Bethe-ansatz method \cite{cTG-BA}. From that solution, one obtains the
ground-state energy density (in physical units),
\begin{equation}
\varepsilon (\rho _{T})=\frac{\pi ^{2}\hbar ^{2}}{6m}\rho _{T}^{2},  \notag
\end{equation}%
with $\rho _{T}=\sum_{j=1,2}|\phi _{j}|^{2}$ being the total density of the
mixture. In terms of the density-functional theory with the local-density
approximation, this amounts to the consideration of the following energy
functional:
\begin{equation}
E[\rho _{T}]=\int_{-\infty }^{+\infty }dx\left\{ \sum_{j=1}^{2}\phi
_{j}^{\ast }\left[ -\frac{\hbar ^{2}}{2m}\frac{d^{2}}{dx^{2}}\right] \phi
_{j}+\rho _{T}\varepsilon (\rho _{T})\right\} ,  \label{E}
\end{equation}%
where chemical potentials $\mu _{j}$ are introduced as Lagrangian
multipliers to guarantee the conservation of the numbers of atoms in the two
species. In the case of equal masses and \emph{fully symmetric} interactions,%
\begin{equation}
\alpha _{1}=\alpha _{2}=\chi =\pi ^{2}/\left( 2m\right) ,  \label{equal}
\end{equation}
Eq. (\ref{coupledTG}), i.e., in this case,
\begin{gather}
\tilde{\mu}_{j}\phi _{j}=-\frac{d^{2}\phi _{j}}{dx^{2}}+  \notag \\
\pi ^{2}(|\phi _{j}|^{4}+|\phi _{3-j}|^{4}+2|\phi _{j}|^{2}|\phi
_{3-j}|^{2})\phi _{j},\;\;j=1,2,  \label{TG-symm}
\end{gather}%
coincide with the equations which provide for the minimization of energy
functional (\ref{E}).

In the following we investigate DW and BD solutions of Eqs. (\ref%
{coupledNLSE}), both for the fully symmetric interaction strengths and
relatively small deviations from this special case.
\begin{figure}[tbph]
\includegraphics[width=0.35\textwidth]{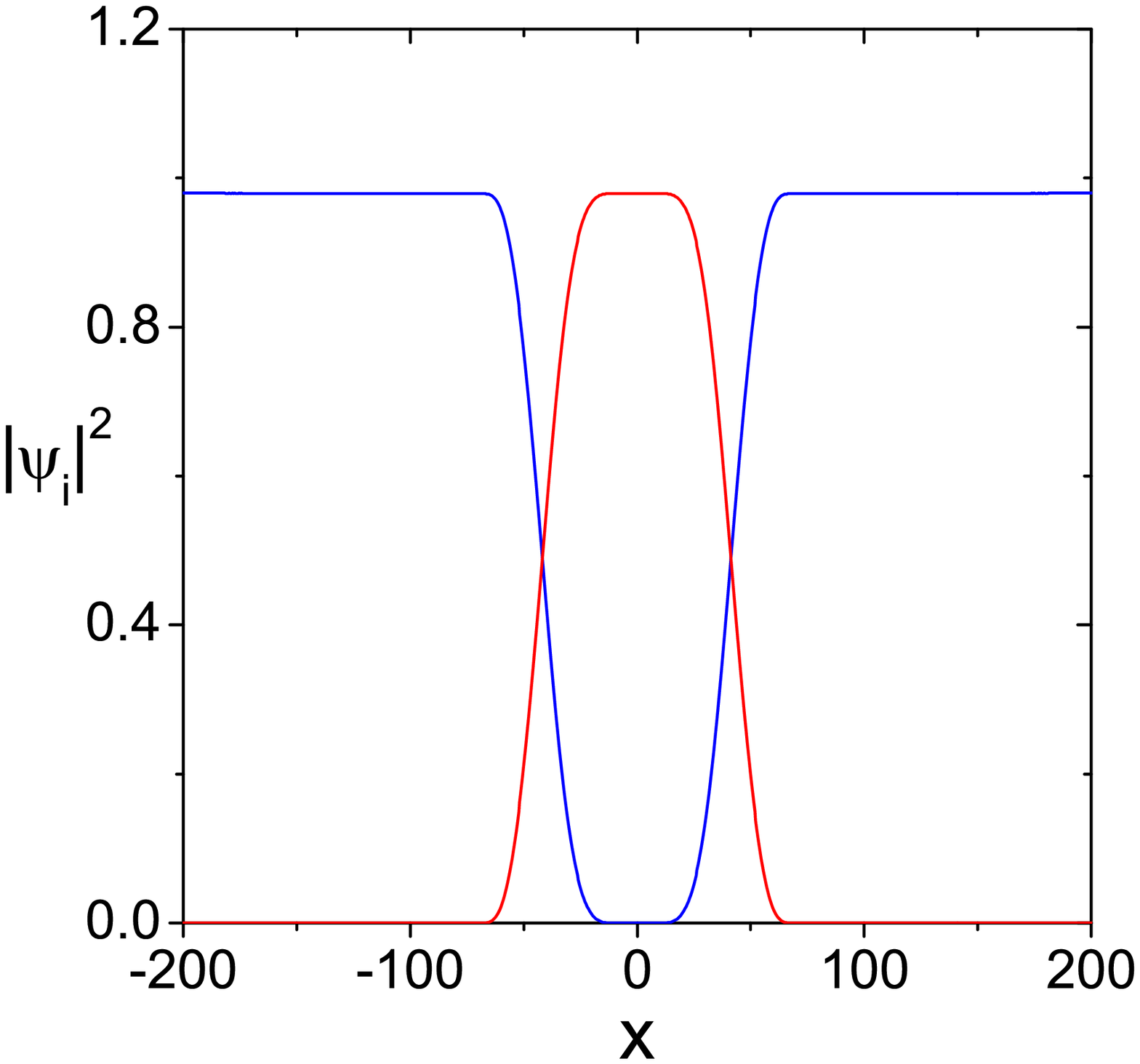} \vskip -3.6cm %
\includegraphics[width=0.35\textwidth]{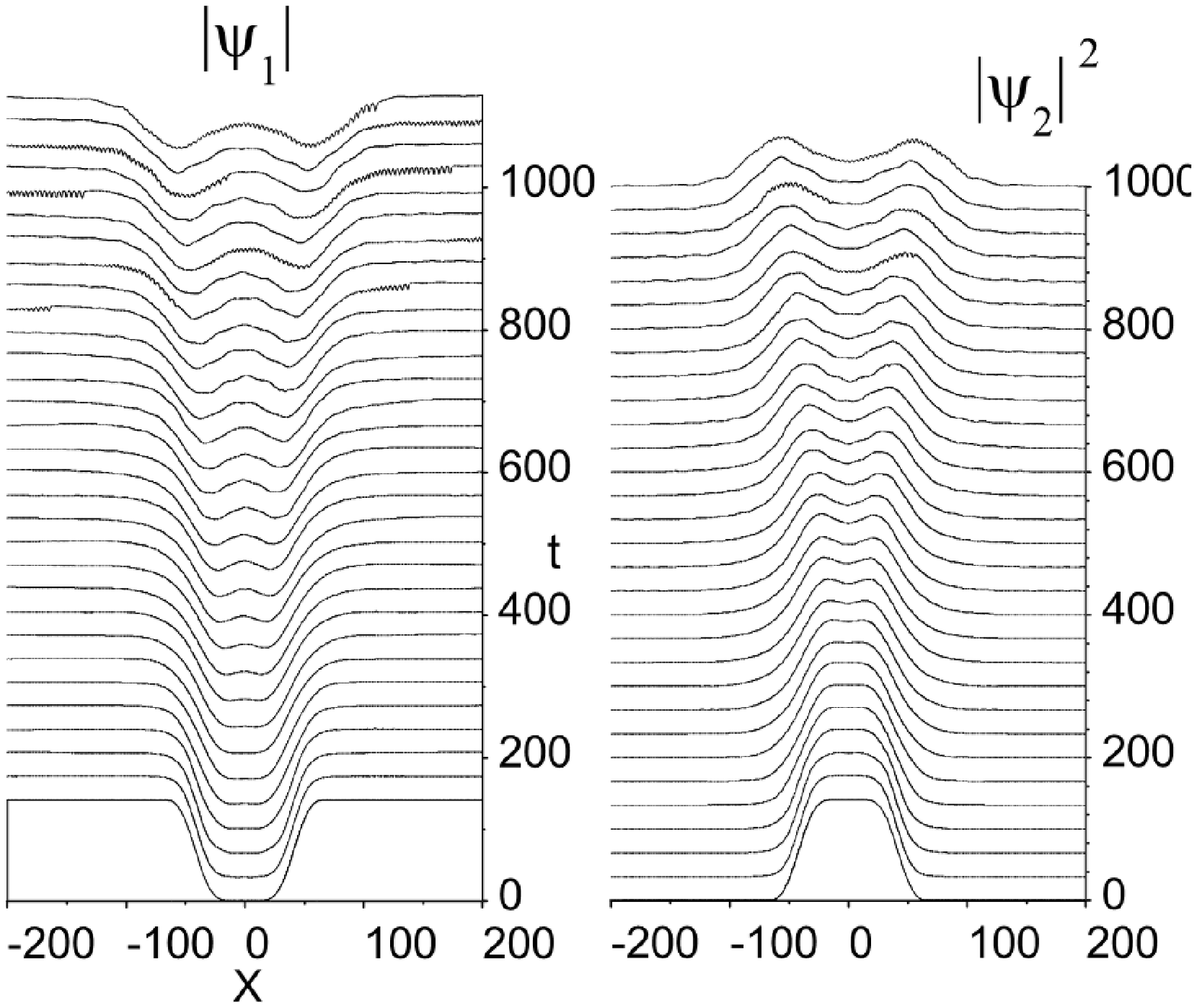} \vskip -3.cm
\caption{(Color online) Top panel: Density profiles of a bubble-drop state,
produced by Eq. (\protect\ref{TG-symm}) as a DW-anti-DW complex, in the
binary TG gas with equal masses and fully symmetric interactions [see Eq. (%
\protect\ref{equal})], i.e., exactly at the miscibility threshold. Bottom
panel: The stability test by real-time simulations of the configuration
depicted in the top panel. The density profiles remain stationary up to $%
t\simeq 400$. Subsequently, they slowly decay into the uniformly mixed
background with oscillations.}
\label{fig7}
\end{figure}

\subsection{The variational approach}

DWs of Eq. (\ref{coupledTG}) with $m_{1}=m_{2}\equiv m,\alpha _{1}=\alpha
_{2}\equiv \alpha ,\;\chi _{1}=\chi _{2}\equiv \chi $, but $\alpha \neq \chi
$, can be studied by means of a variational approximation (VA) based on the
corresponding Lagrangian,
\begin{eqnarray}
L &=&\frac{1}{2}\int_{-\infty }^{+\infty }\left\{ \left[ \frac{1}{2}\left(
\frac{d\phi _{1}}{dx}\right) ^{2}+\left( \frac{d\phi _{2}}{dx}\right) ^{2}%
\right] -\mu \left( \phi _{1}^{2}+\phi _{2}^{2}\right) \right.  \notag \\
&&\left. +\frac{\alpha }{3}\left( \phi _{1}^{6}+\phi _{2}^{6}\right) +\chi
\left( \phi _{1}^{4}\phi _{2}^{2}+\phi _{1}^{2}\phi _{2}^{4}\right) +\frac{2%
}{3}\mu ^{3/2}\right\} dx,  \notag \\
\label{L}
\end{eqnarray}%
%
where constant $\left( 2/3\right) \mu ^{3/2}$ is the\textit{\ counter-term},
which is added to cancel the divergence of the integral at $\left\vert
x\right\vert \rightarrow \infty $. The DW solution may be approximated by an
ansatz similar to the one adopted for the BEC mixture in Eq. (\ref{ansatz}),
i.e.,
\begin{equation}
\phi _{1}(x)=\sqrt{\frac{\sqrt{\mu }}{1+e^{\lambda x}}},\;\;\phi _{2}(x)=%
\sqrt{\frac{\sqrt{\mu }}{1+e^{-\lambda x}}},  \label{ans}
\end{equation}%
with $\lambda $ considered as a free variational parameter. The substitution
of Eq. (\ref{ans}) into Lagrangian (\ref{L}) yields the corresponding
effective Lagrangian,
\begin{equation}
L_{\mathrm{eff}}=\frac{\sqrt{\mu }}{2}\left[ \frac{\lambda }{8}+\left( \chi
-\alpha \right) \frac{\mu }{\lambda }\right] .  \label{Leff}
\end{equation}%
Finally, the variational equation, $\partial L_{\mathrm{eff}}/\partial
\lambda =0$, produces the main result of the VA,
\begin{equation}
\lambda ^{2}=8\mu \left( \chi -\alpha \right) .  \label{VA}
\end{equation}%
This analysis suggest that, in the TG mixtures with equal masses and fully
symmetric interactions ($\chi =\alpha $), DW states cannot exist, i.e., they
should degenerate into a mixed uniform background, as one can see from Eq. (%
\ref{VA}). This prediction correlates with the fact that the immiscibility
threshold in Eq. (\ref{immC}), in the absence of cubic interactions and for
equal asymptotic densities, reduces to $\alpha =\chi $, hence the TG mixture
with the fully symmetric interactions sits precisely at the miscibility
threshold.
\begin{figure}[tbph]
\centerline{
\includegraphics[width=0.23\textwidth]{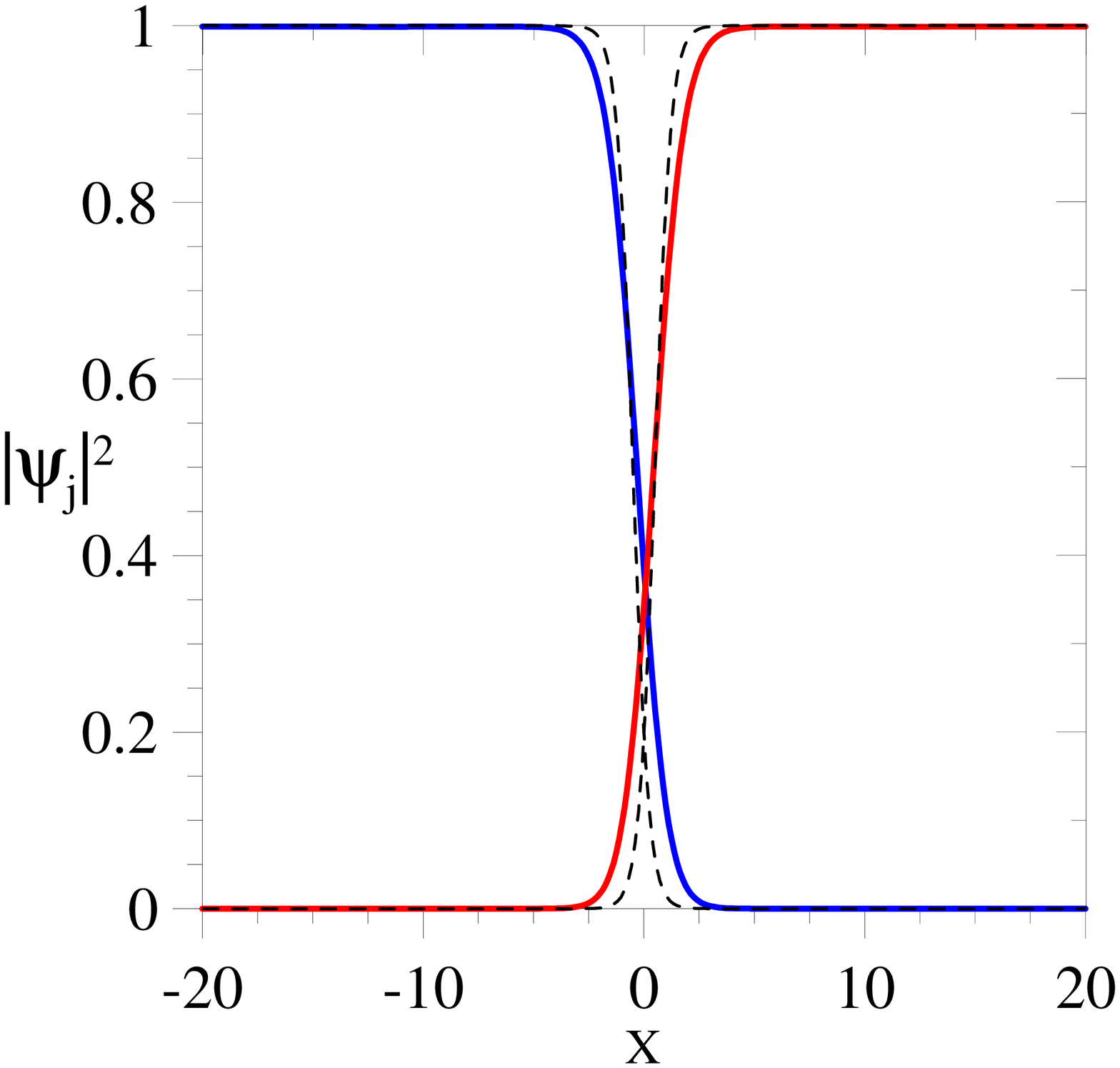}
\includegraphics[width=0.23\textwidth]{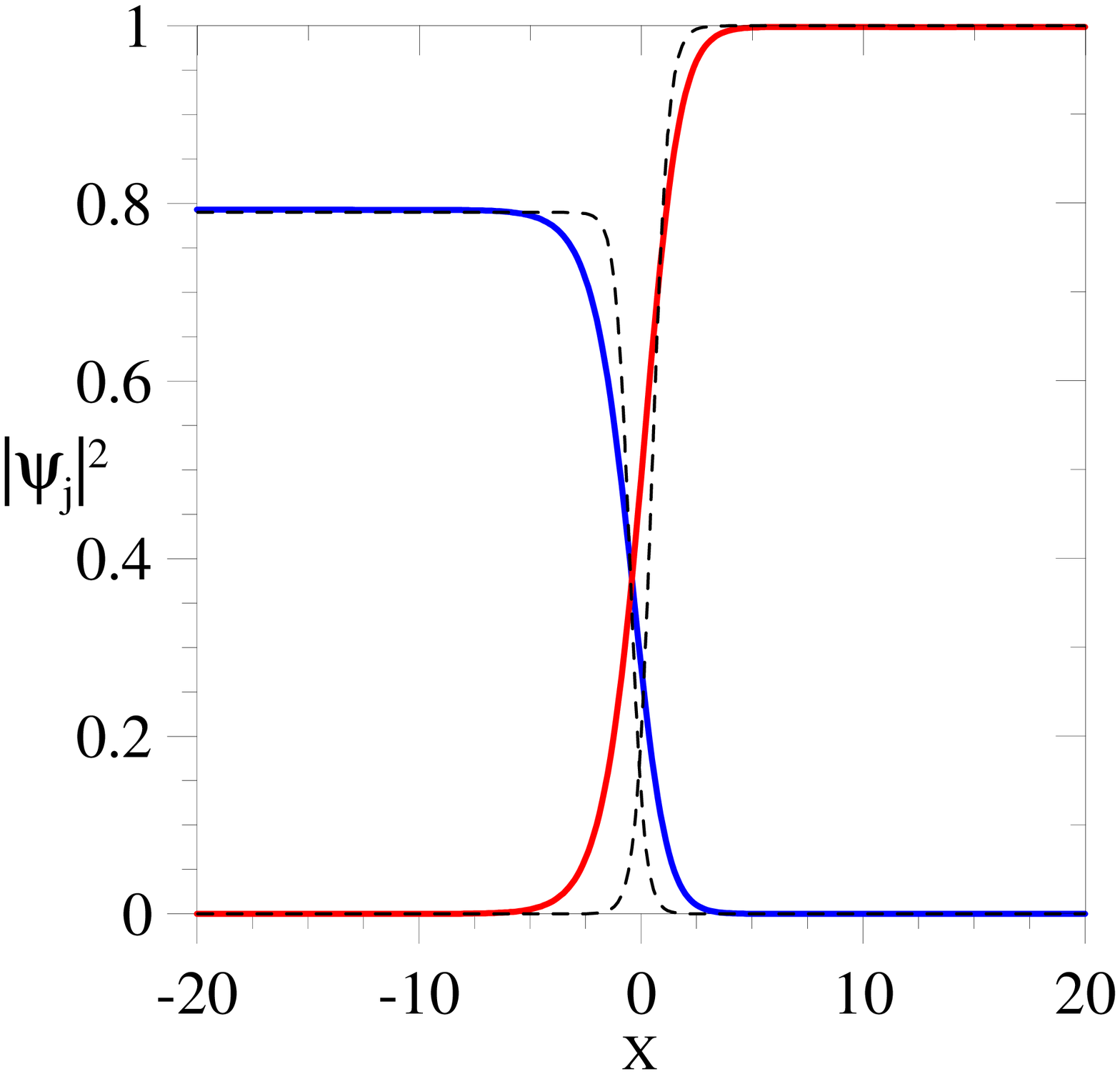}
}
\caption{(Color online) DW density profiles produced by Eq. (\protect\ref%
{coupledTG}) for equal masses $m_{1}=m_{2}=1$ and inter-species interaction
strength $\protect\chi =2$. Other parameters are fixed as $\protect\alpha %
_{1}=\protect\alpha _{2}=1$ (left panel) and $\protect\alpha _{1}=2$, $%
\protect\alpha _{2}=1$ (right panel). Blue and red lines refer to first and
second components, respectively, while the black dashed lines denote VA
results.}
\label{profile}
\end{figure}

The numerical analysis of the model based on Eq. (\ref{TG-symm}) with the
full interaction symmetry show that DW and BD patterns indeed decay into
uniformly mixed backgrounds. An example of this is shown in Fig. \ref{fig7},
where a stationary numerical BD solution and its perturbed evolution are
depicted, respectively, in the top and bottom panels. As seen, such a
stationary state exists at the miscibility threshold, but it turns out to be
unstable. In this connection, it is relevant to mention that, although for
the TG mixture the time evolution governed by the quintic NLSE may have no
direct physical meaning, it can be used as a mathematical tool to verify if
a stationary DW or BD state is well defined, and also to check if the
mixture indeed sits at the miscibility threshold. This can be done, as
usual, by adding a small random perturbation to the initial stationary
profiles and letting the state evolve according to the time-dependent
quintic NLSEs associated with stationary equations (\ref{TG-symm}). The
bottom panel of Fig. \ref{fig7} shows that, although the BD remains
stationary for a relatively long time ($t\approx 400$), still later the
profiles slowly decay into the uniform mixed background state. This behavior
is consistent with the fact that the TG-TG mixture finds itself precisely at
the miscibility threshold, as predicted by the above analysis.

However, the simulations demonstrate that, as soon as one deviates from the
fully-symmetric-interaction limit defined by Eq. (\ref{equal}), the DW and
BD states become well-defined ground states, depending on the b.c. This
point is further investigated in the next section.


\subsection{The asymmetric setting}

The results of the previous section can be extended for the binary TG gas
with asymmetric inter-species and intra- species interaction strengths
and/or unequal masses, provided that the deviation from the fully symmetric
setting is not too large. As the corresponding quantum problem is not
exactly solvable in the absence of the full symmetry, such a model is
substantiated less rigorously than the full-symmetry limit.

In the following, we assume an asymmetric system with different coefficients
of the quintic self-interaction in Eq. (\ref{coupledNLSE}), $\alpha _{1}\neq
\alpha _{2}$. The accordingly modified system obviously keeps Lagrangian (%
\ref{L}), making it possible to perform a similar VA as in the previous
section. Due to the asymmetry, however, the DW solution should be looked for
with unequal chemical potentials, $\mu _{1}\neq \mu _{2}$. Then, the DW
solution is defined by the following b.c.:%
\begin{eqnarray}
\phi _{1}^{2}\left( x=-\infty \right) &=&\sqrt{\mu _{1}/\alpha _{1}},~\phi
_{2}^{2}\left( x=-\infty \right) =0,  \notag \\
&&  \label{bc} \\
\phi _{1}^{2}\left( x=+\infty \right) &=&0,~\phi _{2}^{2}\left( x=+\infty
\right) =\sqrt{\mu _{2}/\alpha _{2}}.  \notag
\end{eqnarray}%
It is worth to mention here that direct substitution demonstrates that the
simplest ansatz based on Eq. (\ref{ansatz}) does not produce any exact DW
solution of Eq. (\ref{coupledNLSE}). Nevertheless, it is possible to find an
exact relation between $\mu _{1}$ and $\mu _{2}$ which is necessary for the
existence of the DW solution. Indeed, Eqs. (\ref{coupledNLSE}) may be
formally considered as equations of motion for a mechanical system with two
degrees of freedom, $\phi _{1}(x)$ and $\phi _{2}(x)$, where $x$ play the
role of formal time. The Hamiltonian of this mechanical system is
\begin{eqnarray}
&&H_{\mathrm{mech}}=\sum_{n=1,2}\Bigg[\frac{1}{4m_{n}}\left( \frac{d\phi _{n}%
}{dx}\right) ^{2}  \notag \\
&&+\frac{1}{2}\mu _{n}\phi _{n}^{2}-\frac{\alpha _{n}}{6}\phi _{n}^{6}-\frac{%
\chi }{2}\phi _{n}^{2}\phi _{3-n}^{4}\Bigg].  \label{H}
\end{eqnarray}%
Because $H_{\mathrm{mech}}$ must take identical values at $x=\pm \infty $,
b.c. (\ref{bc}) demonstrate that $\mu _{1}$ and $\mu _{2}$ are related by
\begin{equation}
\alpha _{2}\mu _{1}^{3}=\alpha _{1}\mu _{2}^{3}.  \label{rel}
\end{equation}%
By means of obvious rescalings, we can eventually set
\begin{equation}
\alpha _{2}=m_{2}=\mu _{2}\equiv 1,  \label{1}
\end{equation}%
and Eq. (\ref{rel}) then yields, for the DW solution, the single admissible
value of the chemical potential of species $\phi _{1}$:
\begin{equation}
\mu _{1}=\alpha _{1}^{1/3}.  \label{mu2}
\end{equation}%
Thus there remain three free parameters: coefficient $\chi $ of the
inter-species quintic interaction, along with $m_{1}$ and $\alpha _{1}$.

We have checked by means of numerical methods that DW solutions exist for a
wide range of values of parameters $\chi ,\alpha _{1,2}$ away from the
fully-symmetric-interaction limit considered in the previous section. Two
typical examples of such numerically found DWs are depicted in Fig. \ref%
{profile} for $\chi \neq \alpha _{1}=\alpha _{2}\equiv \alpha $ (the left
panel) and $\chi =\alpha _{1}\neq \alpha _{2}$ (the right panel). Note that
in the former (resp. latter) case the DW components have equal (resp.
unequal) backgrounds, as a consequence of the equality (resp. inequality) of
the intra-species interaction strengths. In the same figure are also
depicted, by dashed lines, the predictions of the VA for the two cases,
which demonstrate reasonable agreement with the numerical findings.

In addition to the DW, a physically relevant state may be the BD complex
similar to those studied above for the BEC mixture. 
In the general case, the BD is a spatially even
solution, with $\phi _{1,2}(-x)=\phi _{1,2}\left( x\right) $, which obeys
the following b.c.:
\begin{eqnarray}
\phi _{1}^{2}\left( x=+\infty \right) &=&0,~\phi _{2}^{2}\left( x=+\infty
\right) =\sqrt{\mu _{2}/\alpha _{2}},  \label{infty} \\
\phi _{1}^{\prime }\left( x=0\right) &=&\phi _{2}^{\prime }\left( x=0\right)
=0.  \label{0}
\end{eqnarray}%
The extreme case of the BD complex is the one with
\begin{equation}
\phi _{2}(x=0)=0,  \label{extreme}
\end{equation}%
when field $\phi _{1}$ completely ousts $\phi _{2}$ at the central point, $%
x=0$.

The conservation of $H_{\mathrm{mech}}$ along $x$ [see Eq. (\ref{H})] has
its bearing for the BD solution too. Indeed, condition $H_{\mathrm{mech}%
}(x=+\infty )=H_{\mathrm{mech}}(x=0)$ yields the following relation for the
densities of the two species at the central point, $x=0$ [see Eq. (\ref{0}%
)]:
\begin{gather}
\sum_{n=1,2}\left[ \mu _{n}\phi _{n}^{2}(x=0)-\frac{\alpha _{n}}{3}\phi
_{n}^{6}(x=0)\right]  \notag \\
-\chi \phi _{1}^{2}(x=0)\phi _{2}^{2}(x=0)\left[ \phi _{1}^{2}(x=0)+\phi
_{2}^{2}(x=0)\right] =  \notag \\
=\frac{2\mu _{2}^{3/2}}{3\sqrt{\alpha _{2}}}.  \label{x=0}
\end{gather}%
This cumbersome relation simplifies for the extreme type of the BD defined
above by condition (\ref{extreme}):%
\begin{equation}
\mu _{1}\phi _{1}^{2}(x=0)-\frac{\alpha _{1}}{3}\phi _{1}^{6}(x=0)=\frac{%
2\mu _{2}^{3/2}}{3\sqrt{\alpha _{2}}}.  \label{simple}
\end{equation}%
The same normalization (\ref{1}) as used above for the DW may be adopted for
the BD solutions. Then, the general family of such solutions depends on four
independent parameters: $\chi $, $m_{1}$, $\alpha _{1}$, and $\mu _{1}$ [the
latter constant is no longer determined by an additional condition, such as
Eq. (\ref{mu2})].


\section{Discussion and conclusions}

In this work, we have introduced the concept of DWs (domain walls) in the two-component TG (Tonks-Girardeau) gas, as well as in the binary BEC described by coupled GPEs (Gross-Pitaevskii equations) with CQ (cubic-quintic) nonlinearities. 
Thus, we have extended the concept of DW previously elaborated for immiscible binary BEC, described by the system of two coupled GPEs  with cubic terms.

In several cases, exact or approximate analytical solutions have been found.
In particular, for the binary BEC with the CQ nonlinearity, exact solutions
for DWs were obtained in the symmetric system, with equal intra-species
scattering lengths. For the existence of these solutions, the presence of
the quintic interactions is necessary. Numerical analysis was used to prove
the existence and stability of asymmetric DWs in the BEC mixtures with
asymmetric CQ interactions in the two components.

In addition to the DWs, we have also investigated DB\ (droplet-bubble)
complexes in the same settings, which consist of a dark (gray) soliton in
one component (the ``bubble"), and a bright soliton (the
``droplet") in the other. In the BEC system, the DWs and DB
states are mobile, keeping their shape in the state of motion.

We have also introduced symmetric and asymmetric DWs in the binary TG gas,
for which the system of two coupled NLSEs (nonlinear Schr\"{o}dinger
equations) with quintic-only repulsive terms was adopted. In particular, we
have showed that a TG mixture with equal atomic masses and fully symmetric
interactions [as per Eq. (\ref{equal})] sits precisely at the immiscibility
threshold, therefore DWs and BDs exist only as metastable states in this
case. The VA (variational approximation) and numerical analysis, however,
demonstrate the existence of stable DWs ground states at a small deviation
from the full symmetry.

Stability of DWs and BD complexes in BEC mixtures with symmetric and
asymmetric CQ interactions in the two components has also been demonstrated.
In this respect, we have derived a general condition for the immiscibility,
which is valid for both the BEC and TG settings. The stability of the DW and
BD states is secured if the immiscibility condition is satisfied.

It is relevant to discuss possible implementations of the above results in
experimental settings. DWs and anti-DWs can be observed (either being
separated, or merged into BD complexes) in toroidal quasi-one-dimensional
traps for gases with scattering lengths tuned so as to satisfy the
immiscibility condition. Note that toroidal traps are routinely created in
laboratories with the aid of magnetic fields \cite{toroidal}, and, as said
above, the scattering length can be easily changed via the
Feshbach-resonance technique. To create BD complexes in the ring, one can
load the gases into two different semicircles of the trap, initially kept
separated by laser sheets, which are slowly removed after the loading.

\section*{Acknowledgements}

MS acknowledges partial support from the Ministero dell' Istruzione, dell'
Universit\`{a} e della Ricerca (MIUR) through a \textit{Programma di Ricerca
Scientifica di Rilevante Interesse Nazionale} (PRIN) 2010-2011 initiative.

GF acknowledges partial financial support from PON Ricerca e Competitivit\`a
2007-2013 under grant agreement PON NAFASSY, PONa3\_00007.

\end{document}